\documentclass[fleqn,12pt]{wlscirep}
\usepackage[utf8]{inputenc}
\usepackage[T1]{fontenc}
\usepackage[numbers,super]{natbib}
\usepackage{caption}

\newcommand{\OIII}{O\,{\scriptsize III}}
\newcommand{\OII}{O\,{\scriptsize II}}

\newcommand{\NIII}{N\,{\scriptsize III}}
\newcommand{\CIII}{C\,{\scriptsize III}}

\newcommand{\NV}{N\,{\scriptsize V}}

\newcommand{\HeII}{He\,{\scriptsize II}}

\graphicspath{{./}{figures/}}

\title{The nature of an ultra-faint galaxy in the cosmic Dark Ages seen with \textit{JWST}}

\author[1,*]{Guido Roberts-Borsani}
\author[1]{Tommaso Treu}
\author[2]{Wenlei Chen}
\author[3]{Takahiro Morishita}
\author[4]{Eros Vanzella}
\author[5]{Adi Zitrin}
\author[4,6]{Pietro Bergamini}
\author[7]{Marco Castellano}
\author[7]{Adriano Fontana}
\author[8]{Karl Glazebrook}
\author[6,9]{Claudio Grillo}
\author[2]{Patrick L. Kelly}
\author[7]{Emiliano Merlin}
\author[8]{Themiya Nanayakkara}
\author[7]{Diego Paris}
\author[4,10]{Piero Rosati}
\author[11]{Lilan Yang}
\author[6,9]{Ana Acebron}
\author[7,12]{Andrea Bonchi}
\author[13,14]{Kit Boyett}
\author[15,16]{Maru\v{s}a Brada\v{c}}
\author[17,18]{Gabriel Brammer}
\author[19,20,21]{Tom Broadhurst}
\author[7]{Antonello Calabr\'o}
\author[22]{Jose M. Diego}
\author[23]{Alan Dressler}
\author[5]{Lukas J. Furtak}
\author[24]{Alexei V. Filippenko}
\author[25,26]{Alaina Henry}
\author[25]{Anton M. Koekemoer}
\author[27]{Nicha Leethochawalit}
\author[1]{Matthew A. Malkan}
\author[17,18]{Charlotte Mason}
\author[28,29]{Amata Mercurio}
\author[1,13,14]{Benjamin Metha}
\author[7]{Laura Pentericci}
\author[25]{Justin Pierel}
\author[2]{Steven Rieck}
\author[26]{Namrata Roy}
\author[7]{Paola Santini}
\author[17,18]{Victoria Strait}
\author[30]{Robert Strausbaugh}
\author[13,14]{Michele Trenti}
\author[31]{Benedetta Vulcani}
\author[32]{Lifan Wang}
\author[33,34,35]{Xin Wang}
\author[36]{Rogier A. Windhorst}

\affil[1]{Department of Physics and Astronomy, University of California, Los Angeles, CA 90095, USA}
\affil[2]{School of Physics and Astronomy, University of Minnesota, 116 Church Street SE, Minneapolis, MN 55455, USA}
\affil[3]{Infrared Processing and Analysis Center, Caltech, 1200 E. California Blvd., Pasadena, CA 91125, USA}
\affil[4]{INAF - OAS, Osservatorio di Astrofisica e Scienza dello Spazio di Bologna, via Gobetti 93/3, I-40129 Bologna, Italy}
\affil[5]{Physics Department, Ben-Gurion University of the Negev, P.O. Box 653, Be'er-Sheva 84105, Israel}
\affil[6]{Dipartimento di Fisica, Universit\`a degli Studi di Milano, Via Celoria 16, I-20133 Milano, Italy}
\affil[7]{INAF Osservatorio Astronomico di Roma, Via Frascati 33, 00078 Monteporzio Catone, Rome, Italy}
\affil[8]{Centre for Astrophysics and Supercomputing, Swinburne University of Technology, PO Box 218, Hawthorn, VIC 3122, Australia}
\affil[9]{INAF - IASF Milano, via A. Corti 12, I-20133 Milano, Italy}
\affil[10]{Dipartimento di Fisica e Scienze della Terra, Universit\`a degli Studi di Ferrara, Via Saragat 1, I-44122 Ferrara, Italy}
\affil[11]{Kavli Institute for the Physics and Mathematics of the Universe, The University of Tokyo, Kashiwa, Japan 277-8583}
\affil[12]{ASI-Space Science Data Center, Via del Politecnico, I-00133 Roma, Italy}
\affil[13]{School of Physics, University of Melbourne, Parkville 3010, VIC, Australia}
\affil[14]{ARC Centre of Excellence for All Sky Astrophysics in 3 Dimensions (ASTRO 3D), Australia}
\affil[15]{University of Ljubljana, Department of Mathematics and Physics, Jadranska ulica 19, SI-1000 Ljubljana, Slovenia}
\affil[16]{Department of Physics and Astronomy, University of California Davis, 1 Shields Avenue, Davis, CA 95616, USA}
\affil[17]{Cosmic Dawn Center (DAWN)}
\affil[18]{Niels Bohr Institute, University of Copenhagen, Jagtvej 128, 2200 København N, Denmark}
\affil[19]{Department of Theoretical Physics, University of the Basque Country UPV/EHU, Bilbao, Spain}
\affil[20]{Donostia International Physics Center (DIPC), 20018 Donostia, Spain}
\affil[21]{IKERBASQUE, Basque Foundation for Science, Bilbao, Spain}
\affil[22]{Instituto de F\'isica de Cantabria (CSIC-UC). Avda. Los Castros s/n. 39005 Santander, Spain}
\affil[23]{The Observatories, The Carnegie Institution for Science, 813 Santa Barbara St., Pasadena, CA 91101, USA}
\affil[24]{Department of Astronomy, University of California, Berkeley, CA 94720-3411, USA}
\affil[25]{Space Telescope Science Institute, 3700 San Martin Dr., Baltimore, MD 21218, USA}
\affil[26]{Center for Astrophysical Sciences, Department of Physics \& Astronomy, Johns Hopkins University, Baltimore, MD 21218, USA}
\affil[27]{National Astronomical Research Institute of Thailand (NARIT), Mae Rim, Chiang Mai, 50180, Thailand}
\affil[28]{Dipartimento di Fisica ``E.R. Caianiello'', Universit\`a degli Studi di Salerno, Via Giovanni Paolo II 132, I-84084 Fisciano (SA), Italy}
\affil[29]{INAF-Osservatorio Astronomico di Capodimonte, Via Moiariello 16, 80131 Napoli, Italy}
\affil[30]{Minnesota Institute For Astrophysics, University of Minnesota, 106 Pleasant St SE, Minneapolis, MN 55455, USA}
\affil[31]{INAF Osservatorio Astronomico di Padova, vicolo dell’Osservatorio 5, 35122 Padova, Italy}
\affil[32]{Mitchell Institute for Fundamental Physics \& Astronomy, Texas A\&M University, Department of Physics and Astronomy, 4242 TAMU, College Station, TX 77843, USA}
\affil[33]{School of Astronomy and Space Science, University of Chinese Academy of Sciences (UCAS), Beijing 100049, China}
\affil[34]{National Astronomical Observatories, Chinese Academy of Sciences, Beijing 100101, China}
\affil[35]{Institute for Frontiers in Astronomy and Astrophysics, Beijing Normal University,  Beijing 102206, China}
\affil[36]{School of Earth and Space Exploration, Arizona State University, Tempe, AZ 85287-1404, USA}

\affil[*]{Corresponding author}

\makeatletter
\renewcommand{\@maketitle}{%
{%
\thispagestyle{empty}%
\vskip-36pt%
{\raggedright\sffamily\bfseries\fontsize{20}{25}\selectfont \@title\par}%
\vskip10pt
{\raggedright\sffamily\fontsize{12}{16}\selectfont  \@author\par}
\vskip25pt%
}%
}%
\makeatother

\begin{document}

\flushbottom
\maketitle

\noindent\textbf{In the first billion years after the Big Bang, sources of ultraviolet (UV) photons are believed to have ionized intergalactic hydrogen, rendering the Universe transparent to UV radiation. Galaxies brighter than the characteristic luminosity $L^{*}$\cite{robertson15, stark16} do not provide enough ionizing photons to drive this cosmic reionization. Fainter galaxies are thought to dominate the photon budget; however they are surrounded by neutral gas that prevents the escape of the Lyman-$\alpha$ photons, which has been the dominant way to identify them so far. JD1 was previously identified as a triply-imaged galaxy with a magnification factor of 13 provided by the foreground cluster Abell 2744\cite{zitrin14}, and a photometric redshift of $z\sim10$. Here we report the spectroscopic confirmation of this very low luminosity ($\sim0.05 L^{*}$) galaxy at $z=9.79$, observed 480 Myr after the Big Bang, by means of the identification of the Lyman break and redward continuum, as well as multiple $\gtrsim4\sigma$ emission lines, with the Near-InfraRed Spectrograph (NIRSpec) and Near-InfraRed Camera (NIRCam) instruments. The combination of the \textit{James Webb Space Telescope} (\textit{JWST}) and gravitational lensing shows that this ultra-faint galaxy ($M_{\rm UV}=-17.35$) -- with a luminosity typical of the sources responsible for cosmic reionization -- has a compact ($\sim$150 pc) and complex morphology, low stellar mass (10$^{7.19}$ M$_\odot$), and subsolar ($\sim$0.6 $Z_{\odot}$) gas-phase metallicity.
} \\

\begin{figure*}[!htb]
\center
 \includegraphics[width=0.675\columnwidth]
 {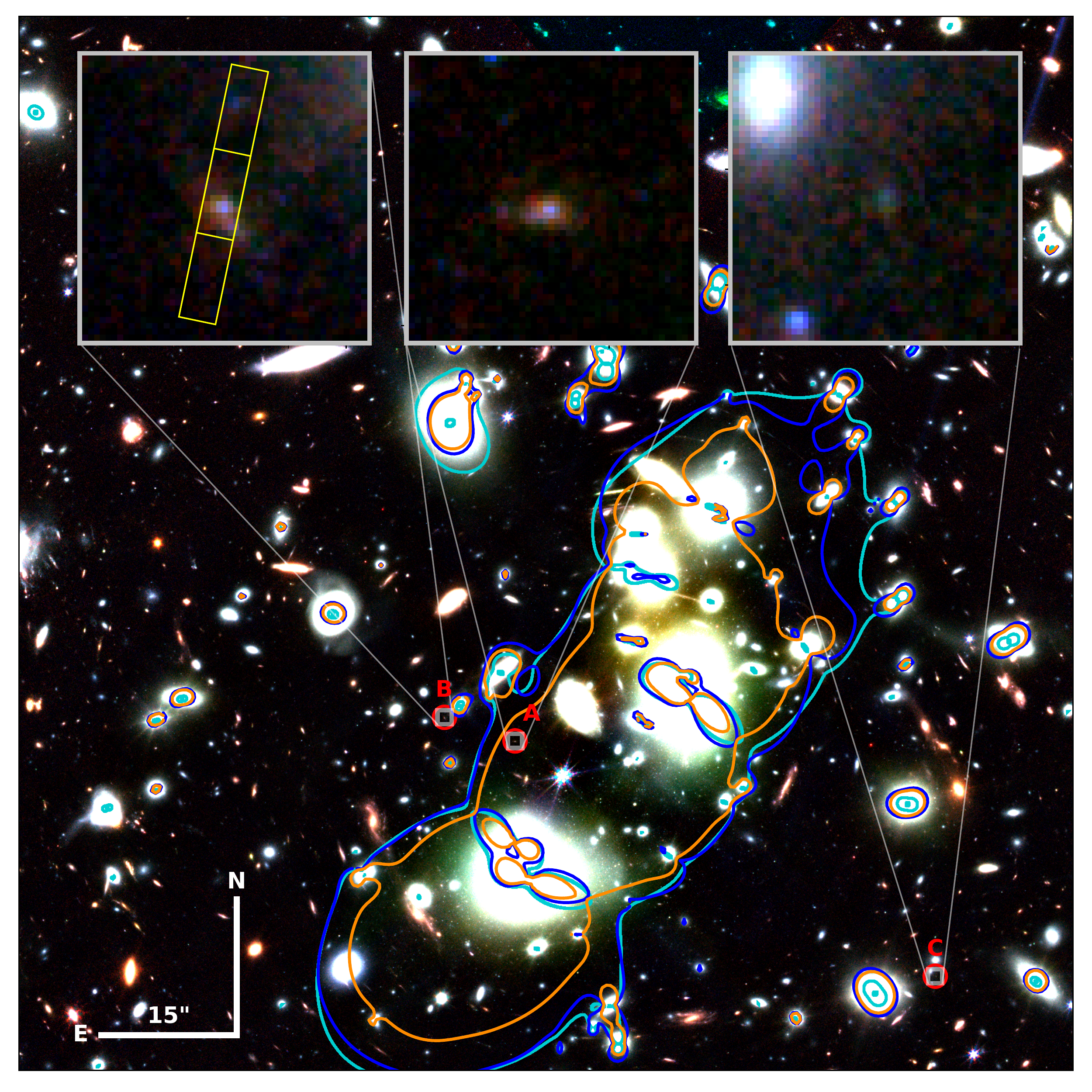} \\
 {\small \caption{\textbf{A false-colour NIRCam image of the Abell 2744 cluster}. Critical curves of formally infinite magnification from two lensing models ([\!\!\citenum{bergamini22}] and an update to [\!\!\citenum{zitrin14}], discussed in the Methods section) are plotted as blue and cyan curves ([\!\!\citenum{bergamini22}] in blue and the [\!\!\citenum{zitrin14}] update in cyan). The A, B and C multiple images of the $z=9.79$ galaxy are highlighted by red circles. The inset plots represent zoom-ins on each of the triply-imaged components, with the brightest component (JD1) highlighted by the $0.''2\times1.''2$ NIRSpec slit (constructed using three MSA shutters) used for prism spectroscopy in our DDT programme.  Critical lines for a source at $z\sim10$, the redshift indicated by the lens model, the photometry and the spectroscopy, are in excellent mutual agreement -- highlighting the robustness of the lens model -- and fall between the two images consistent with their similar brightness. To show the power of lensing to constrain the redshift of the source, we also show, as an orange line, the critical line for a source at $z=2.24$\cite{bergamini22}. If JD1 were at this lower redshift, the image positions and fluxes would be inconsistent with the measurements. Here B=F115W+F150W, G=F200W+F277W, R=F356W+F444W.} \label{fig:RGB}}
\end{figure*}

The galaxy JD1 was observed by the \textit{JWST} with deep NIRCam imaging (GO 2561; PI Labb\'e and DDT 2756; PI Chen) and NIRSpec prism spectroscopy (DDT 2756; PI Chen). Residing behind the Abell 2744 galaxy cluster, the source is gravitationally lensed and displays three images, of which two bright components (A and B) reside to the north of the main galaxy cluster and a fainter component (C) to the south. The NIRCam images cover all three components, whereas NIRSpec prism spectroscopy targeted the brightest component (component B\footnote{Spectral overlap prevented us from observing both.}) on 23 October 2022. An RGB image using the novel NIRCam data is shown in Figure~\ref{fig:RGB}, where we highlight a portion of the cluster field (including the positions of each lensed image of JD1), with inset plots focused on each component and the spectroscopic target (henceforth JD1). \\

\begin{figure}
\center
 \includegraphics[width=0.8825\columnwidth]{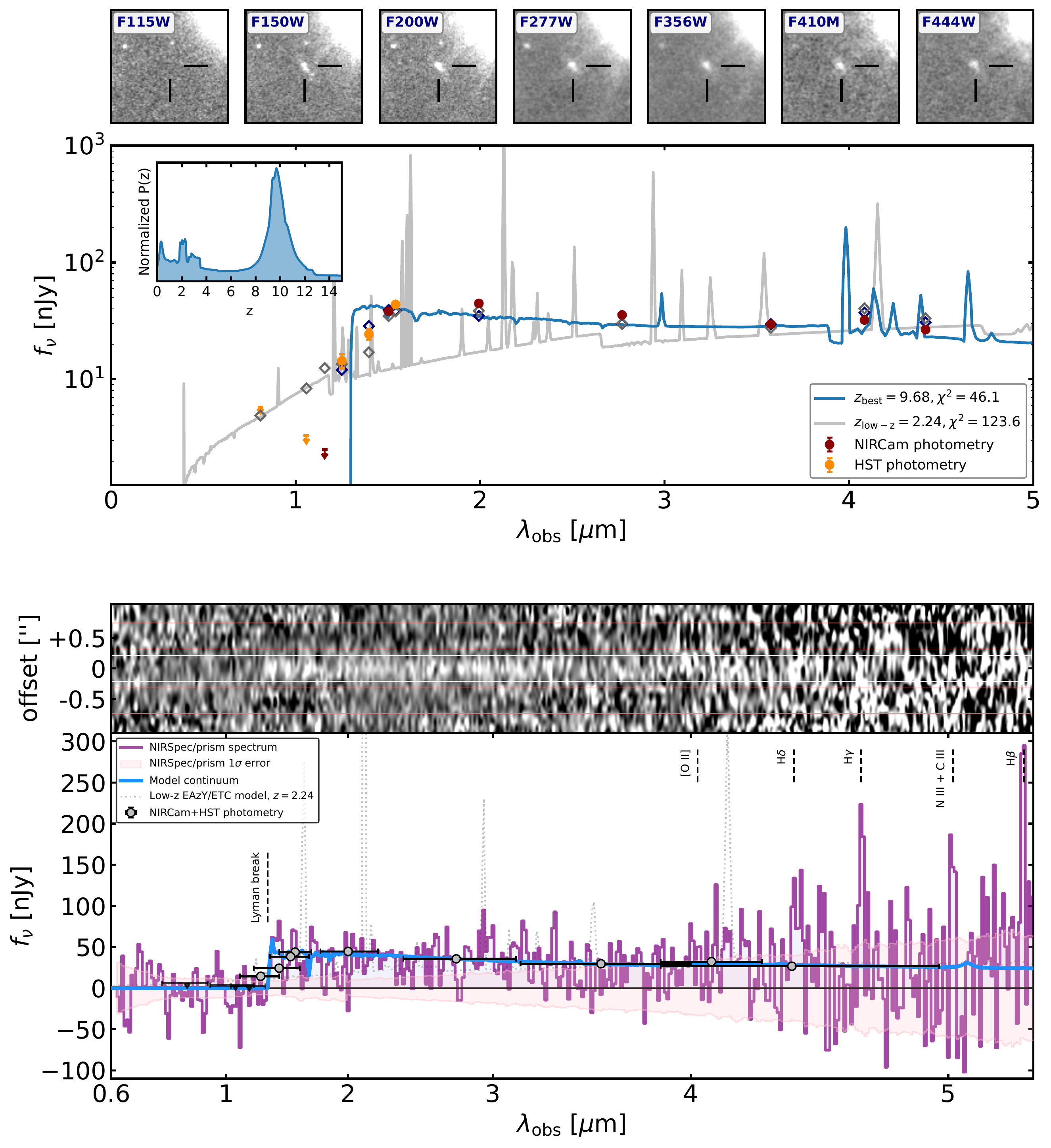}
 {\small \caption{\textbf{The SED and NIRSpec spectrum of JD1.} \textit{Top:} NIRCam postage stamp images (top panels) for each of the NIRCam bands used in this study, along with their extracted photometry (red points, and \textit{HST} photometry in orange), and the best-fit SED (bottom panel, blue curve and diamonds) derived with \texttt{EAzY}. A forced low redshift fit is also shown (grey curve and diamonds) and is disfavored. An inset plot highlights the $P(z)$ of the fit across the entire allowed redshift range. \textit{Bottom:} The 2D (top) and 1D (bottom) NIRSpec prism spectrum of JD1, with the positive and negative 2D traces indicated by white and red lines, respectively (with widths corresponding to the 1D extraction kernel). The optimally-extracted 1D spectrum is shown in purple (line and fill), with associated flux uncertainties (pink fill) and best-fit continuum model in blue. The combined NIRCam and \textit{HST} photometry are shown as grey points, with associated uncertainties in black. Simulated NIRSpec spectra using the \texttt{EAzY} best-fit SEDs are shown as dotted grey and blue lines (low-$z$ and high-$z$, respectively), for illustration. The 1D spectrum is normalized to the F150W band. All uncertainties refer to $1\sigma$ standard deviation.
 }\label{fig:results}}
\end{figure}

The multiband NIRCam photometry (F115W, F150W, F200W, F277W, F356W, F410M, F444W) alone confirms JD1 as a $z\sim10$ photometric candidate. Its colours are consistent with the Lyman break residing between the F115W and F150W filter, where all photons more energetic than rest-frame Lyman-$\alpha$ have been absorbed and scattered by intervening intergalactic hydrogen, while the galaxy is detected at high significance at all wavelengths redward of Lyman-$\alpha$ (that is, the F200W, F277W, F356W, F410M, and F444W filters). Combining the NIRCam data with deep \textit{Hubble Space Telescope} (\textit{HST}) Frontier Field photometry (including WFC3 F160W, F140W, F125W, F105W, and ACS F814W filters) from [\!\!\citenum{zitrin14}], a fit to the spectral energy distribution (SED) with the photometric redshift code, \texttt{EAzY} \cite{brammer08}, yields a precise photometric redshift $z=9.68^{+0.11}_{-0.10}$ (consistent with the geometric redshift predicted from lens models \cite{zitrin14,bergamini22}, $z>8.6$ at 95\% C.L.). The results of the fitting, along with the NIRCam images, are shown in the top panel of Figure~\ref{fig:results}, and include a forced fit of a $z<4$ interloper, for comparison (see also Extended Data Figure \ref{fig:pdf} for a comparison of P(z) constraints resulting from different subsets of the data). A $z\sim10$ solution is clearly favored, with a $P(z)$ governed by a dominant $z_{\rm phot}=9.68$ peak and a smaller, secondary peak at $z_{\rm phot}=2.24$, the latter of which is statistically rejected due to its increased $\chi^{2}$ statistic ($\chi^{2}=46.1$ and $\chi^{2}=123.6$ for the high-$z$ and low-$z$ fits, respectively). The colors of images A and B are identical within the uncertainties, and from a geometric standpoint their morphologies are consistent with the expected lensing configuration, further confirming their identification as multiple images of the same source. A low redshift solution is also ruled out by the positions and comparable fluxes of the multiple images: for a $z\sim10$ solution both images are expected to show similar offsets relative to the critical magnification line, and thus to have comparable fluxes. This contrasts with a $z=2.24$ solution, which would place the critical line far closer to the A component (Figure~\ref{fig:RGB}) and result in a different position.\\

The NIRSpec spectrum of the galaxy (bottom panel of Figure~\ref{fig:results}), spanning a wavelength range $\lambda_{\rm obs}\simeq0.6-5.3 \mu$m, provides the third and conclusive piece of evidence for the redshift identification. A clear continuum break is apparent at $\sim1.3 \mu$m, with continued continuum emission redward of the break and noise (centered at $\sim$0 nJy) blueward of it. Identifying the break as the Lyman break, this alone places the galaxy at a redshift of $z\sim9.75$, in excellent agreement with the photometric redshift estimated above, and deep into the heart of the cosmic Dark Ages when the universe was mostly filled with neutral hydrogen\cite{mason19}. Although [\OIII]$\lambda\lambda$4960,5008 \AA\ emission lines are redshifted out of the NIRSpec/prism coverage, we identify a number of marginal emission lines in the rest-frame optical, most notably (but not limited to) the Balmer H$\gamma$ and H$\beta$ lines, which are detected at central wavelengths of 46,856.0 \AA\ and 52,482.1 \AA\ with peak signal-to-noise (S/N) ratios of $4.9\sigma$ and $4.2\sigma$ using independent Gaussian fits, respectively. The two lines (and other marginal line emission) are discussed in the Methods section and refine the redshift of JD1 to $z_{\rm spec}=9.793\pm0.002$. As expected, strong Lyman-$\alpha$ is also not seen near the Lyman break, highlighting the likely consequence of damping-wing scattering by a highly neutral medium and indicating that this galaxy does not reside in a massive, ionized bubble. Adopting the upper limit on flux at the expected position of Lyman-$\alpha$, the wavelength resolution and the F150W photometry as continuum, we set a $2\sigma$ upper limit on the rest-frame equivalent width (EW$_{0}$) of $<9$ \AA. In similar fashion, we find EW$_{0}$=50$\pm$14 \AA\ and EW$_{0}$=68$\pm$15 \AA\ for H$\gamma$ and H$\beta$, respectively, and EW$_{\rm 0}=$1-83 \AA\ (1-28 \AA\ at $\lambda_{\rm obs}<4 \mu$m and 1-83 \AA\ at $\lambda_{\rm obs}>4 \mu$m) for the rest of the NIRSpec spectrum using the best-fit continuum model presented in Section 6 of Methods and adopting wavelength widths set by the native wavelength grid. While dust could play some role, the absence of strong lines in JD1 is likely to be due to a combination of its low overall star formation rate (SFR) (see below), its intrinsic faintness and its magnification. This is consistent with spectroscopic characterizations of Lyman-break galaxies at $z\sim3$ (e.g., [\!\!\citenum{shapley03}]) and both lensed and unlensed galaxies at $z>7$ ([\!\!\citenum{stark17,jiang21,Williams22}), all of show enhanced specific SFRs compared to JD1.\\

\begin{figure}
\center
 \includegraphics[width=\columnwidth]{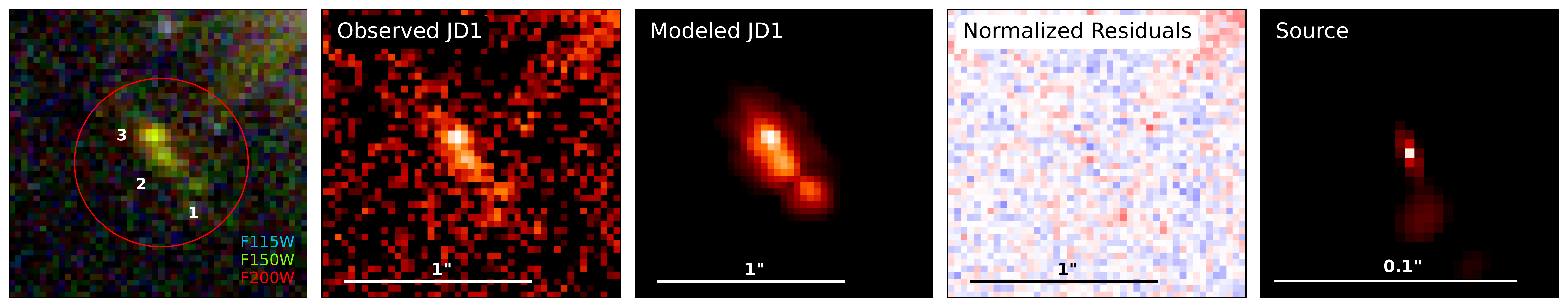}
  {\small \caption{\textbf{The morphology of JD1 from \textit{JWST} NIRCam imaging}. From the left to right: an RGB (F115W, F150W, F200W) image of the galaxy system, the F150W image of the source, the \texttt{lenstruction} model of the source, the ($1\sigma$ flux-normalized) residuals between the F150W data and the model, and the reconstructed source-plane galaxy. The sizes of the cutouts are labelled in each panel.}\label{fig:morphology}}
\end{figure}

\begin{figure}
\center
 \includegraphics[width=0.65\columnwidth]{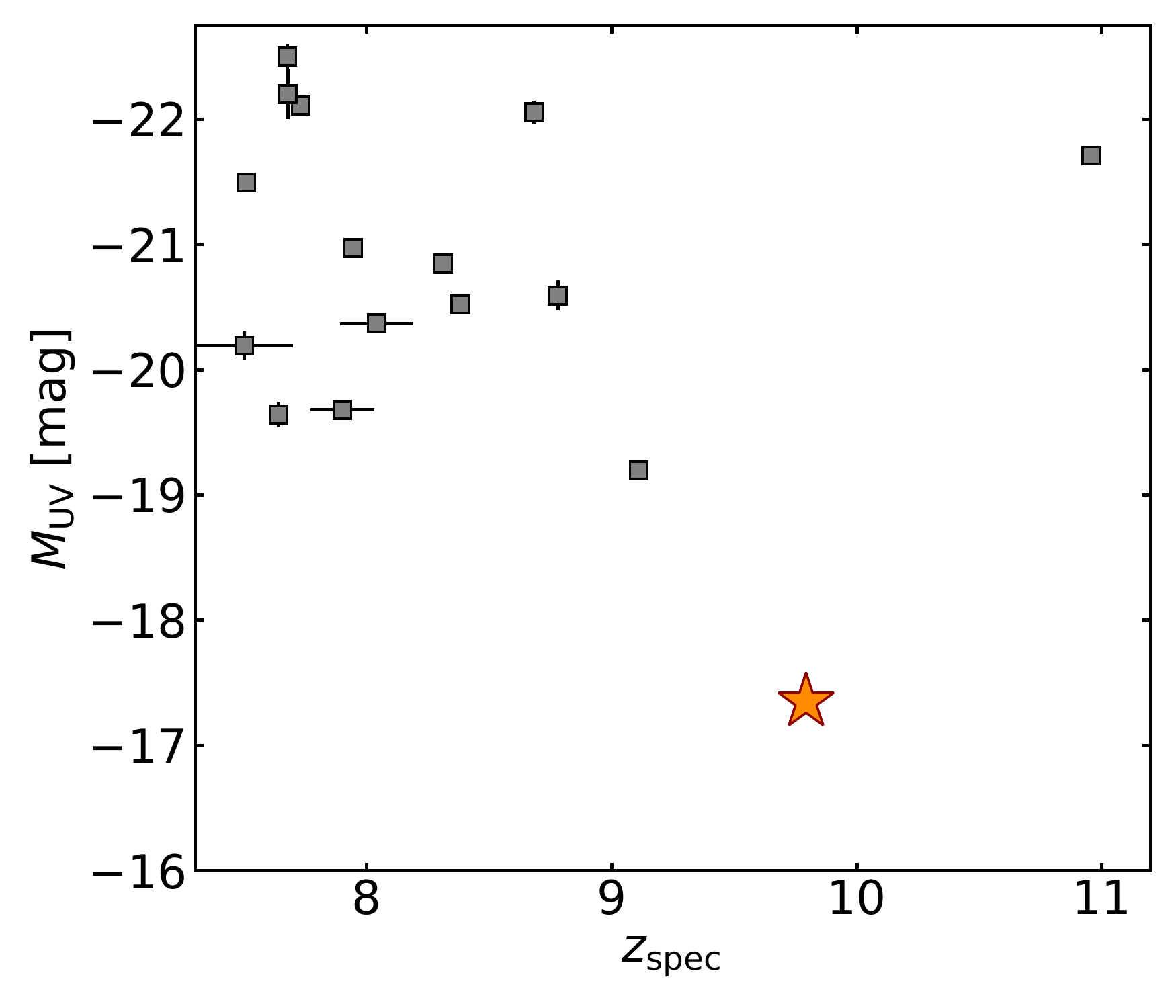} \\
 {\small \caption{\textbf{Spectroscopically confirmed sources and their absolute magnitudes.
 } A compilation of sources in the literature\cite{finkelstein13,oesch15,watson15,zitrin15,oesch16,hoag17,laporte17,hashimoto18,tamura19,bouwens21_rebels,laporte21,topping21,rb22_niriss} with spectroscopic redshifts (from NIR observations of Lyman-$\alpha$ or the Lyman break, or sub-mm observations of FIR lines such as [\OIII] 88 microns) and their associated $M_{\rm UV}$ values (grey squares), compared to JD1 (orange star). For consistency, all literature $M_{\rm UV}$ values are calculated using the sources' $H_{\rm 160}$ photometry and their spectroscopic redshifts, and assuming a UV slope $\beta=-2$. JD1 is by far the lowest luminosity source (by $\sim2$ mag) at comparable redshifts.}\label{fig:MUV}}
\end{figure}

We infer global galaxy properties from detailed SED fitting of the source, incorporating all of the spectroscopic and photometric constraints. Correcting for a fiducial magnification factor of $\mu=13.1^{+0.7}_{-0.7}$\,\cite{bergamini22}, the best-fit model (bottom panel of Figure~\ref{fig:results}) paints a picture of a young ($\sim$30 Myr), star-forming (log SFR/M$_{*}$\,yr$^{-1}=-8.38$) and low stellar mass (log $M_{*}/M_{\odot}=$7.48) galaxy that is dust-poor ($A_{\rm V}=0.20$ mag), subsolar in metallicity (log $Z_{*}/Z_{\odot}=-0.23$) and intrinsically compact (see Figure \ref{fig:morphology} and analysis below). The inferred UV slope of the spectrum ($\beta=-1.90$) is blue and supports such a hypothesis, where the galaxy is dominated by a young, star-forming system that is beginning its chemical enrichment journey. The absolute UV magnitude of the system, $M_{\rm UV }=-$17.44 mag, classifies the galaxy as a sub-$L^{*}$ ($\sim0.05 L^{*}$, adopting $M^{*}_{\rm UV}=-20.6$ mag\cite{bouwens15,oesch18}) system and, given its extreme distance, ranks it as the faintest known source at comparable redshifts (c.f. [\!\!\citenum{oesch16,hashimoto18}]; see Figure~\ref{fig:MUV}). As such, the galaxy luminosity is similar to that  of the sources that are thought to provide the bulk of the UV photons required to reionize the Universe \cite{robertson15,stark16}. We list the best-fit properties and uncertainties in the bottom half of Extended Data Table~\ref{tab:photo}. \\

Finally, the combination of lensing magnification and \textit{JWST}’s extraordinary angular resolution results in an effective source-plane resolution of $\sim$80 pc, allowing us to characterize the morphology of the galaxy (Figure~\ref{fig:morphology}). JD1 comprises two main components: a primary knot (component 3) that likely dominates the signal in the prism spectrum, and an extended component characterized by two fainter knots (components 1 and 2). The complex morphology is also observed in individual NIRCam bands out to $\sim$ 5 $\mu$m (Figure~\ref{fig:results}). The nucleic profiles (components 1 and 3) are well described by a S\'ersic index (component 3) and a pure point spread function (PSF), whereas the more extended nature of component 2 is modelled by an exponential S\'ersic profile (see details in the Methods section). The profiles yield effective radii of $\sim$22-40 mas for the nucleic components and 0.12$''$ for the extended component, which, after de-lensing along the magnification stretch of the source, translate to $\sim18-30$ pc and $100\pm25$ pc, respectively. Using the secure spectroscopic redshift, we adopt a forward-modelling approach with the \texttt{lenstruction} code and deflection maps from [\!\!\citenum{bergamini22}] to reconstruct JD1 in the source-plane: from the NIRCam F150W image (second panel of Figure~\ref{fig:morphology}), we find that an exponential S\'ersic profile combined with two-dimensional (2D) shapelets provides an excellent fit to the data (third and fourth panels of Figure~\ref{fig:morphology}) and results in a compact source-plane morphology with an effective radius of $\sim34$ pc (last panel of Figure~\ref{fig:morphology}). \\

The combination of JD1’s spectroscopic confirmation and magnification affords a unique and unprecedented insight into the physics of an ultra-faint galaxy in the cosmic Dark Ages, providing a first glimpse at the power of JWST and highlighting exactly what the observatory was built for. The unveiling of entire populations of faint galaxies and their physical properties through unbiased measurements now represents the logical next step and leap in our ability to characterize the sources that reionized the Universe.

\newpage

\section*{Figures}

\newpage

\section*{Methods}

\textbf{1. Cosmological model.} The cosmological parameters adopted in this paper are $H_{\rm 0} = 70$ km s$^{-1}$ Mpc$^{-1}$, $\Omega_{\rm m} = 0.3$, and $\Omega_{\rm \Lambda} = 0.7$. All magnitudes are quoted in the AB system\cite{oke83}. \\

\noindent \textbf{2. \textit{JWST}/NIRCam data reduction and photometry.}
We reduced all broadband NIRCam images (F115W, F150W, F200W, F277W, F356W, and F444W) adopting the procedures and methods outlined by [\!\!\citenum{merlin22}]; we refer the reader to that paper for full details but provide a summary here. We used the public Space Telescope Science Institute (STScI) pipeline (v11.16.14) with the latest set of reference files and zeropoints (jwst\_1023.pmap), and began the data reduction with the \texttt{calwebb\_detector1} and \texttt{calwebb\_image2} routines on the uncalibrated raw images to apply flat-field and dark current corrections, and to obtain data quality flags that denote bad pixels or detector level defects. Using a combination of the resulting pixel masks and a customized version of the \texttt{SExtractor} code \cite{bertin96}, further custom procedures were made to the images to correct for persisting $1/f$ noise, ``snowballs'' and stray light. Single exposure images are then stacked in mosaics and re-binned onto a common pixel grid using \textsc{Swarp} \cite{bertin10}. The images were aligned to Gaia DR3 data with the code \textsc{Scamp} \cite{bertin06}, using pre-existing catalogs of Magellan data. The average difference between the two astrometric solutions is of $\sim1$mas, with a normalized median absolute deviation $\sim$15mas. 

The detection image of the galaxy is the co-added F444W image, and extraction of the fluxes was performed in the same manner as in [\!\!\citenum{merlin22}]. In short, we used the code \textsc{a-phot} \cite{merlin19} to measure the fluxes within fixed circular apertures with diameter 2$\times$FWHM of the F444W image ($0.28''$), on PSF-matched images; these fluxes provide robust colors, which were used to scale the total F444W flux (estimated with a Kron elliptical aperture) obtaining total fluxes in all bands. As the object is located near a bright source, it is likely contaminated by the neighbour's light; we therefore applied a local background subtraction module. We find good consistency comparing the total F150W flux with the flux given by [\!\!\citenum{zitrin14}] for WFC3 F160W. The resulting photometry and coordinates of JD1 are listed in the top half of Extended Data Table \ref{tab:photo}.\\

\noindent \textbf{3. \textit{JWST}/NIRSpec observations, data reduction and spectral extraction.}
JD1 is composed of three image components, two of which (components A and B in the north; see Figure~\ref{fig:RGB}) reside sufficiently close to each other to cause spectral overlap and a third that, while residing on the south side of the foreground cluster, appears far fainter than its counterparts. As such, we target the brightest of the three images, component B. The observations were carried out on 23 October 2022 with low-resolution ($R=\lambda/\Delta\lambda\simeq50-350$) prism spectroscopy, adopting a three-shutter slitlet (approximately $0.''2\times1.''2$) in the Micro-Shutter Array (MSA) and nodding pattern over a total exposure time of $\sim1.23$ hrs. The position(s) of the slit (shutters) relative to the galaxy are shown in the inset image of Figure~\ref{fig:RGB}. The observations were carried out over two different pointings but with identical setups and position angles. 

All exposures were reduced using the latest reference files (jwst\_1023.pmap, including in-flight NIRSpec flats), beginning first with the \texttt{calwebb\_detector1} on the raw exposures to correct for detector-level artefacts and converting to count-rate images. We then use the \texttt{msaexp} software for the rest of the data reduction, which includes preprocessing scripts to identify and correct for $1/f$ noise, ``snowballs'', and bias current for each exposure and official STScI pipeline routines for the bulk of the data reduction. Those routines include \texttt{AssignWcs}, \texttt{Extract2dStep}, \texttt{FlatFieldStep}, \texttt{PathLossStep}, and \texttt{PhotomStep}, which perform WCS registration, 2D spectral extraction, flat-fielding, pathloss corrections and both wavelength and flux calibrations on each exposure. The reduced 2D exposures are then background-subtracted using their paired exposures from the three-shutter dither and drizzled to a common grid, before being combined into a single 2D spectrum via the median of their inverse-variance-weighted exposures, including the rejection of outliers. The drizzling and stacking of the NIRSpec exposures allows us to increase the resulting S/N of the spectrum; although this may introduce some correlation between pixels, we limited such effects by adopting a grid at the spectrum's native resolution. All exposures are visually inspected by two authors (GRB and WC) after each main step. One-dimensional (1D) spectra were subsequently extracted from the final 2D image using an optimal extraction procedure\cite{horne86}, which uses the collapsed spectrum (summed over the dispersion direction) to construct a Gaussian extraction kernel based on the spatial profile of the spectrum. The collapse of the spectrum over the full wavelength range ensures that PSF broadening is accounted for in the extraction kernel. The routine is run on the 2D flux image and on the variance image (squaring the weighting factor) to extract the spectrum and 1$\sigma$ uncertainties, the latter of which also incorporates bad pixel masking. The resulting spectrum (and uncertainties) was normalized to the NIRCam/F150W photometry and is presented in the bottom panel of Figure~\ref{fig:results}. 
The wavelength range covers $\lambda_{\rm obs}=0.6-5.3 \mu$m which, for a $z\sim10$ object includes the entire rest-frame UV spectrum, while the flux calibration accuracy across the full wavelength coverage is estimated at 5\% or lower\footnote{\url{https://jwst-crds.stsci.edu/display_all_contexts/}}.
\\

\noindent \textbf{4. Photometric redshift estimates and ETC simulations.}
We used the photometric redshift code, \texttt{EAzY}, to determine a precise photometric redshift using both NIRCam and \textit{HST} photometry (the latter from [\!\!\citenum{zitrin14}]). The \textit{HST} filters used (in addition to the NIRCam bands described above) were WFC3 F160W, F140W, F125W, F105W, and ACS F814W. Although deep \textit{Spitzer}/InfraRed Array Camera (IRAC) data also exist over the Abell 2744 cluster, the relatively poor spatial resolution makes photometry close to the cluster centre uncertain and prone to contamination. As such, we opted not to use those here. We adopted the default set of galaxy SED templates (v1.3), which, among others, include SEDs of dusty, $z\sim2$ star-forming galaxies and templates with strong emission lines. We allowed a redshift range of $z=0-15$ (adopting a flat prior) and fitted a linear combination of the templates to the photometry. All other default settings were adopted. We ran the code on the photometry described above, in addition to two other iterations, one with \textit{HST} photometry only and one with NIRCam photometry only, to assess the constraining power of each dataset in deriving a reliable photometric redshift. The resulting $P(z)$ from each of the fits are shown in Extended Data Figure~\ref{fig:pdf}, where clearly photometry longward of $\sim2 \mu$m is necessary to accurately constrain the location of the Lyman break at $z>9$ and exclude low-$z$ solutions (we note that the photo-$z$ analysis of [\!\!\citenum{zitrin14}] employed $>3 \mu$m upper limits from \textit{Spitzer}/IRAC). The best-fit SED is found to confidently lie at a redshift of $z_{\rm phot}=9.68^{+0.11}_{-0.10}$.

For illustration purposes, we use the best-fit SEDs to simulate expected NIRSpec spectra from each, using the \textit{JWST} Exposure Time Calculator (ETC) and adopting our observational setup. The simulations are plotted as dashed grey and blue lines (for the low-$z$ and high-$z$ SEDs, respectively) in Figure~\ref{fig:results} (bottom panel) and highlight the inconsistency between our observed NIRSpec spectrum and the prominent emission lines expected from a low-$z$ solution.\\

\noindent \textbf{5. Strong lensing models.}
To check the robustness of our results with respect to the lens model, we used two lens models, described below. One is the recently published model by [\!\!\citenum{bergamini22}], and the second is a major update of the [\!\!\citenum{zitrin14}] model. The two models are based on a large number of observational constraints and represent the state of the art. They give mutually consistent results, so our conclusions are independent on the choice of lens model. When necessary, we use the magnification from [\!\!\citenum{bergamini22}], but all the inferred numbers would be the same within the uncertainties for the other model. 

In the first model, detailed by [\!\!\citenum{bergamini22}], the cluster total mass distribution was reconstructed using the positions of 90 multiple images from 30 different point-like sources, with spectroscopic redshifts between 1.69 and 5.73. Internal velocity dispersion measurements of 85 cluster galaxies were exploited to calibrate the subhalo mass component, which includes 225 cluster galaxies. For this work, we enhanced this model by including three additional photometric and strongly lensed sources identified with the recently acquired \textit{JWST}/Near-InfraRed Imager and Slitless Spectrograph (NIRISS) data with a total of eight multiple images, including two (A and B) of JD1. We also updated the positions of all the mass components and multiple images in the lens model, adopting the new \textit{JWST} astrometric grid. The total root mean square (r.m.s.) separation between the observed and model-predicted positions is 0.37 arcsec (the same as in the original model [\!\!\citenum{bergamini22}]). To estimate the median magnification values and the CL intervals, we extracted 500 random sets of parameter values from the final Markov chain Monte Carlo (MCMC) chain, which has a total of 2.5 $\times 10^{6}$ samples. The predicted magnification values, with their 1$\sigma$ confidence level statistical uncertainties, are $\mu_{\rm A}=12.0^{+1.0}_{-0.9}$ and $\mu_{\rm B}=13.1^{+0.7}_{-0.7}$ for the multiple images A and B, respectively. The magnification ratio is in excellent agreement with the photometric measurements of images A and B. The redshift value of JD1 predicted by this strong lensing model is $z_{\rm lens}>8.6$ at the 95\% confidence level, assuming a uniform prior between $z=2$ and $z=12$, thereby fully consistent with the spectroscopic and photometric redshifts of image B. 

The second lens model used here is an update to the model for Abell 2744 presented by [\!\!\citenum{zitrin14}], revisited here as follows. As input we started with the list of cluster members, and multiple image systems from [\!\!\citenum{bergamini22}], which are based on recent spectroscopic measurements from the Very Large Telescope's Multi Unit Spectroscopic Explorer (MUSE) instrument. Following [\!\!\citenum{bergamini22}], external mass clumps around the main central cluster core are also included in this update. However, we limited the external clump positions to places where strong lensing supports them, that is, we identify a few potential multiply-imaged systems in the northern part of the cluster, and used them to improve the part of the model for which only weak lensing measurements have thus far been available. Furthermore, as our goal was to estimate properties of the high-redshift source images, we also included images A and B as constraints. We built our updated model with a revised, analytic version of the parametric method used by [\!\!\citenum{zitrin15b}] to model the Cluster Lensing And Supernova survey with Hubble (CLASH) sample. The main update is that the new version is not coupled to a specific grid resolution, and thus can achieve more accurate results. Similar to other parametric methods, the model assumes two main components: a superposition of all clusters members, parametrized each here as double pseudo-isothermal elliptical density profiles, based on common scaling relations \cite{jullo07}; and a dark matter component consisting of larger-scale halos, parametrized here as pseudo-isothermal elliptical mass distributions. In addition, the model can accommodate external shear, if warranted by the data. In addition to the mass associated with the luminous galaxies, we distribute five dark matter halos: two halos are centred on the two central bright galaxies, but their central position is free to move; the other three are fixed to the three bright galaxies in the northern part of the field, about 2.5 arcminutes northwest of the main clump, seen in extended images of the field (e.g. [\!\!\citenum{merten11}]). Following common practice, we adopted a scaling relation to connect the galaxies to their dark matter halos. We left as free parameters the velocity dispersion and cut-off radius of a $L^{*}$ galaxy, and the exponents of the relations themselves. For all dark matter halos, the velocity dispersion, core radius, ellipticity and position angle were free to be optimized. We also left free the velocity dispersion of six bright galaxies in the central part, and the ellipticity parameters of the brightest cluster galaxy (BCG). The minimisation is done in the source plane, via a MCMC procedure. We obtain a r.m.s. value of $\simeq0.6$ arcsec. We obtain magnification estimates of $10.7\pm0.6$, $11.9\pm0.8$, and $3.1\pm0.2$ for images A, B, and C, respectively, consistent with the ones obtained by [\!\!\citenum{bergamini22}]. Further details of the model will be given by Furtak et al. (in prep.). \\

\noindent \textbf{6. Spectrophotometric modelling and analyses.}
We conducted a SED-fitting analysis by using a publicly available code, {\tt gsf}, which allowed us to fit both photometric and spectroscopic data simultaneously\cite{morishita19}. The code generates model templates by using {\tt fsps} \cite{conroy09} with the Initial Mass Function (IMF) set to the parameters of [\!\!\citenum{chabrier03}]. We assume an Small Magellanic Cloud dust attenuation curve \cite{gordon03}. The spectral templates are matched to the resolution of the observed spectrum, $R\sim100$, for spectral data, whereas broadband data points are fitted with the model photometry after convolving the templates by the corresponding filters. We adopted a non-parametric form for star formation histories by following the method presented in [\!\!\citenum{morishita19}]. For the redshift of JD1, we generated multiple templates of different ages (1, 3, 10, 30, 100 and 300 Myr). An emission line component of ionization parameters log\,$U \in [-3:0]$, also generated by {\tt fsps}, was added by the amount of an amplitude parameter. Dust attenuation, metallicity of the stellar templates and redshift were left as free parameters. In sum, we had 6+2+1+1+1=11 free parameters. The posterior distribution of the parameters was sampled by using \texttt{emcee} \cite{emcee13} for $10^{5}$ iterations with the number of walkers set to 100. The final posterior was collected after excluding the first half of the realisations (known as burn-in). The resulting best-fit continuum model is shown in Figure~\ref{fig:results} (bottom panel); the physical parameters quoted in the main text and tabulated in Extended Data Table~\ref{tab:photo} are the 50th percentile of the posterior distribution, along with the 16th to 84th percentile uncertainty ranges. The SFR is calculated as the average of the posterior star formation history over the last 100\,Myrs. \\

\noindent \textbf{7. Source morphology.}
JD1 shows a clear elongated shape along the magnification stretch in which three main components are identified (two nucleated, labelled 1 and 3, and one extended, labelled 2, as shown in Figure~\ref{fig:morphology}). The morphology of the system is analysed by exploiting the \textit{JWST}/NIRCam images presented here. As performed in [\!\!\citenum{vanzella22}] we ran GALFIT\cite{peng10} to reproduce the observed JD1 image in the NIRCam/F150W band. This band offers the sharpest PSF while probing pure stellar continuum not affected by Lyman-$\alpha$ emission (if any) and the attenuation from the intergalactic medium. The underlying extended component (2) is modelled by adopting a S\'ersic light profile ($n=4$), effective radius of 4 pixels, a position angle of 53 deg and an axis ratio ($b/a$) of 0.2. The nucleated emissions were modelled adopting a pure PSF for component 1 and a S\'ersic ($n=1$) model for component 3. Very similar results are obtained adopting Gaussian ($n=0.5$), exponential ($n=1$) or $n=4$ indexes, for both the nucleated (when applying a S\'ersic model) and the extended components. Specifically, the pure PSF model of component 1 corresponds to an upper limit on the radius, which can be associated to the half width at half maximum of the PSF ($\simeq$40 mas, in the observed plane, in the F150W image). The brighter component 3 is well reproduced with a S\'ersic model and effective radius of $\sim$22 mas (approximately 0.7 pixels). The extended component is less constrained, however, with an effective radius of 0.12$''$ (observed). After de-lensing along the magnification stretch (adopting $\mu_{\rm tang} = 5.2$\cite{bergamini22}), such angular scales translate to an upper limit of 30 pc for the compact component 1, of order 18$^{+8}_{-5}$ pc for component 3 and 100$\pm$25 pc for the extended one (component 2). However, it is worth stressing that the size of the elongated component can be underestimated because of the faintness of the system.

We further performed a source-reconstruction of JD1 in the source plane, by following the procedure outlined by [\!\!\citenum{yang20}], using the \texttt{lenstruction} lensing code and a forward-modelling approach. Again using the F150W image, the parameterization of JD1 in the source-plane was done using an elliptical S\'ersic profile to describe the central knot (which likely dominates the NIRSpec spectrum) and 2D shapelets to describe the extended and irregular component. The light profile is then subjected to the deflection maps of the [\!\!\citenum{bergamini22}] lens model and convolved to the F150W PSF to mimick lensing effects and match the surface brightness of JD1. The F150W data and results of the modelling are shown in the second and third panels of Figure~\ref{fig:morphology}, where we find an excellent fit to the data by the model, as seen by the normalized residuals between the two (normalized by the $1\sigma$ data uncertainties; see the fourth panel of Figure~\ref{fig:morphology}). The source-plane model of JD1 (last panel) is found to be very compact, with a half-light radius of approximately 34 pc. \\

\noindent \textbf{8. Identification of marginal emission lines and flux limits.}
We observe flux peaks in both the 1D and 2D spectra that stand out from local fluctuations, at observed wavelengths of approximately $4.68 \mu$m and $5.25 \mu$m. Using the redshift of the Lyman break, the rest-frame wavelengths fall on the expected positions of H$\gamma$ and H$\beta$. We inspect the individual exposures of the NIRSpec spectrum as well as a variety of extraction profile widths to ensure the putative lines are not a result of artefacts (e.g., snowballs, cosmic ray hits or the width of the extraction profile). We see no evidence to suggest they may be and note they are only seen in the combined spectrum, as expected given their flux. We fitted the two lines with simple Gaussian profiles using \texttt{emcee}\cite{emcee13}. Using the peaks of the fits and the standard deviation of the local flux immediately adjacent to the lines, we measured peak S/N ratios of $4.9\sigma$ and $4.2\sigma$ for the putative H$\gamma$ and H$\beta$ lines. The integrated line ratio of the two is H$\gamma$/H$\beta=0.81\pm0.38$, consistent within uncertainties with ratios reported in the literature for Sloan Digital Sky Survey (SDSS) galaxies (0.458-0.475; [\!\!\citenum{groves12}]). The central wavelengths place them at $z=9.792\pm0.004$ and $z=9.793\pm0.001$ (for H$\delta$ and H$\beta$, respectively), in near-perfect agreement with the independent redshift measurement from the Lyman break ($\delta z\simeq0.043$), thus refining the redshift of JD1 to $z_{\rm spec}=9.793\pm0.002$ (taken as the mean redshift of the two lines). Assuming Gaussian probabilities, we find the detection of both of these lines at their significance and at coincidental redshifts to be a 6.6$\sigma$ event. We note that, at a redshift of $z_{\rm spec}=9.793$, strong [\OIII]$\lambda\lambda$4960,5008 \AA\ emission lines are redshifted out of the NIRSpec/prism coverage.

Additionally, the spectrum also reveals a prominent peak at approximately $5.01$ $\mu$m, in perfect agreement with the redshift given by the Balmer lines, and which coincides with the locations of the \NIII$\lambda\lambda\lambda$4634,4640,4641 \AA\ and \CIII$\lambda\lambda\lambda$4647,4650,4651 \AA\ triplets. The line profile (measured as a single Gaussian fit given the low spectral resolution and proximity of the lines between themselves) is sufficiently broad to encompass both triplets, and is thus likely the result of the combined contribution from those six lines. Using our best-fit continuum model, we measure EW$_{0}$=78$\pm$19 \AA. Although such lines are not common among the general galaxy population, their origin derives from dense and energetic stellar winds ejected by massive Wolf-Rayet (WR) stars. Although WR galaxies are comparatively rare among the general population, they have been found in their hundreds at low redshift from rest-frame optical studies with, for example, the SDSS, CALIFA and MaNGA surveys (e.g., [\!\!\citenum{brinchmann08,mc16,liang20}]). The absence of significant flux at the locations of more highly ionized \NV$\lambda\lambda\lambda$3479,3484,4058 \AA\ and \NV$\lambda\lambda$4603,4619 \AA\ lines would indicate WR stars of type WN8h-WN9h, characterized by stellar masses, radii and effective temperatures of $\gtrsim$30 $M_{\odot}$, $\gtrsim$20 $R_{\odot}$, and approximately 35,000-40,000 $K$.\cite{crowther07}

The spectrum of WN8h-WN9h WR stars is also dominated by low ambient metallicity, which could explain (together with the lower spectral resolution at bluer wavelengths) the apparent absence of the [\OII]$\lambda\lambda$3726,3729 \AA\ doublet (EW$_{0}<26$ \AA, at $2\sigma$) and lead to significantly weaker stellar winds that reduce the line widths of the profiles\cite{crowther06a,crowther06b}. With an ionization potential effectively double that of the above triplets (that is, 54.4 eV cf. 29.6 and 24.4 eV, respectively), \HeII$\lambda$4685 \AA\ requires extreme radiation fields more commonly associated with active galactic nuclei\cite{feltre16}. As such, its apparent absence (EW$_{0}<12$ \AA, at $2\sigma$), and the absence of other high ionization lines in the rest-frame UV, is unsurprising and supports the hypothesis of a metal-poor system without extreme radiation fields from active galactic nuclei. As such, one hypothesis is that the gravitational amplification and stretching of JD1's image (Figures 1 and 3) provides -- together with the position of the NIRSpec slit -- a unique viewpoint into a source-plane region of the galaxy where the light is dominated by a population of these massive stars rather than the integrated light dominated by massive OB stars.

No other comparable flux peaks (positive or negative) are seen in the 1D spectrum of JD1. Although there are a very small number of similar peaks in the red ($>4 \mu$m) portion of the 2D spectrum outside of the kernel trace, no other such examples show multiple peaks that simultaneously (i) fall at redshifts coinciding with the Lyman break and (ii) show similar spatial extent to JD1's continuum profile. The detection of the three lines discussed above at their measured significance and at coincidental redshifts is found to be a 7.6$\sigma$ event. Nevertheless, we consider only the marginal Balmer lines sufficiently detected for a minor refining of the galaxy redshift, and ultimately both deeper and higher resolution data will be required to verify the presence and origin of all the tentative emission lines reported here.

\begingroup

\renewcommand{\section}[2]{}%

\endgroup

\section*{Code Availability}
Our analysis makes use of three primary codes, all of which are publicly available. The photometric redshift analyses were performed with \texttt{EAzY}, the latest version of which (including the templates used here) is available at \url{https://eazy-py.readthedocs.io/en/latest/}. The data reduction of the NIRCam images were performed with the official STScI \textit{JWST} pipeline, which can be found here: \url{https://github.com/spacetelescope/jwst}. The NIRSpec data were reduced using the \texttt{msaexp} code, which can be found here: \url{https://github.com/gbrammer/msaexp}. The reduced NIRSpec spectrum was analysed with the \texttt{gsf} code, which is available here: \url{https://github.com/mtakahiro/gsf}. The morphological source-plane reconstruction was done with \texttt{lenstruction}, found here: \url{https://github.com/ylilan/lenstruction}.

\section*{Data Availability}
All data used in this paper are publicly available through the Mikulski Archive for Space Telescopes (MAST) server with the relevant program IDs (2561 for the NIRCam imaging, 2756 for the NIRSpec spectroscopy). All other data generated throughout the analysis are available from the corresponding author upon request or at \url{https://github.com/guidorb/jwst-nirspec-jd1}.

\section*{Acknowledgements} GRB thanks Nicolas Laporte for valuable conversations regarding the NIRSpec spectrum of the source. This work is based on observations made with the NASA/ESA/CSA JWST. The data were obtained from the Mikulski Archive for Space Telescopes at the Space Telescope Science Institute, which is operated by the Association of Universities for Research in Astronomy, Inc., under NASA contract NAS 5-03127 for JWST. These observations are associated with program JWST-ERS-1324. We acknowledge financial support from NASA through grant JWST-ERS-1324. AA, AM, PB, CG, PR and EV acknowledge financial support through grants PRIN-MIUR 2017WSCC32, and 2020SKSTHZ. AA has received funding from the European Union’s Horizon 2020 research and innovation programme under the Marie Sklodowska-Curie grant agreement No 101024195 — ROSEAU. AZ and LF acknowledge support by Grant No. 2020750 from the United States-Israel Binational Science Foundation (BSF) and Grant No. 2109066 from the United States National Science Foundation (NSF), and by the Ministry of Science \& Technology, Israel. CM acknowledges support by the VILLUM FONDEN under grant 37459. The Cosmic Dawn Center (DAWN) is funded by the Danish National Research Foundation under grant DNRF140. BM acknowledges support from an Australian Government Research Training Program (RTP) Scholarship. This research is supported in part by the Australian Research Council Centre of Excellence for All Sky Astrophysics in 3 Dimensions (ASTRO 3D), through project number CE170100013. J.M.D. acknowledges the support of projects PGC2018-101814-B-100 and MDM-2017-0765. A.V.F. is grateful for support from the Christopher R. Redlich Fund and numerous individual donors. XW is supported by CAS Project for Young Scientists in Basic Research, Grant No. YSBR-062. MB acknowledges support from the Slovenian national research agency ARRS through grant N1-0238. RAW acknowledges support from NASA JWST Interdisciplinary Scientist grants NAG5-12460, NNX14AN10G and 80NSSC18K0200 from GSFC. We acknowledge support from the INAF Large Grant 2022 ``Extragalactic Surveys with JWST''  (PI Pentericci).

\section*{Author Contributions}
GRB led the NIRSpec data reduction and analysis with input from several coauthors. GRB and TT wrote the paper, and developed the main interpretation of the results. AF, MC, EM, and DP reduced the NIRCam images and provided the photometry. TM performed the spectral-fitting and derived the physical parameters. AA, CG, PB, PR developed and updated the reference lens model providing predictions on lensed quantities. AM contributed to the construction of the photometric and spectroscopic catalogs for the lens model. AZ and LF constructed the new lens model and wrote the accompanying text. TN, KG, and WC assisted with NIRSpec data reduction and performed quality checks on the 1D and 2D spectra. EV, PR,and LY performed the analysis of the source morphology. All authors discussed the results and commented on the manuscript. 

\section*{Competing interests} The authors declare no competing interests.

\section*{Corresponding Author.} Guido Roberts-Borsani (guidorb@astro.ucla.edu)

\section*{Extended Data}

\captionsetup[table]{name=Extended Data Table}

\begin{table}[ht]
\begin{center}
\begin{tabular}{lc}
\hline \noalign {\smallskip}
Property & Value    \\
\hline 
\hline
RA (J2000) [deg] & 3.5950014\\
Dec (J2000) [deg] & -30.4007411\\
$\mu$ & 13.1$^{+0.6}_{-0.7}$\\
F115W [AB] & $<$29.79 ($2\sigma$) \\
F150W [AB] & 27.44$\pm$0.05 \\
F200W [AB] & 27.28$\pm$0.04 \\
F277W [AB] & 27.52$\pm$0.04 \\
F356W [AB] & 27.72$\pm$0.04 \\
F410M [AB] & 27.63$\pm$0.07 \\
F444W [AB] & 27.84$\pm$0.06 \\
$z_{\rm EAzY}$ & 9.68$^{+0.11}_{-0.10}$ \\
\hline
$z_{\rm spec}$ & 9.793$\pm$0.002\\
log $M_{*}$ [$M_{\odot}$] & 7.19$^{+0.17}_{-0.12}$\\
log $Z_{*}$ [$Z_{\odot}$] & -0.20$^{+0.14}_{-0.50}$\\
log $T_{*}$ [Gyr] & -1.87$^{+0.77}_{-0.53}$\\
$A_{\rm v}$ [mag] & 0.55$^{+0.11}_{-0.06}$\\
log SFR [$M_{*}$/yr] & -1.14$^{+0.14}_{-0.10}$\\
$\beta_{\rm UV}$ & $-$2.22$^{+0.03}_{-0.02}$\\
$M_{\rm UV}$ & $-$17.35$^{+0.03}_{-0.03}$\\
\hline
\end{tabular}
\vspace{-0.3cm}
{\small \caption{
\textbf{Top: a summary of the NIRCam photometry and \texttt{EAzY} redshift of JD1.} All fluxes are \textit{uncorrected} for magnification, and the fiducial magnification factor is from [\!\!\citenum{bergamini22}]. \textbf{Bottom: the best-fit global properties of JD1.} The properties derive from the SED analysis presented in the Methods section. All values are \textit{corrected} for magnification where needed. All uncertainties refer to $1\sigma$ standard deviation unless stated otherwise.
\label{tab:photo}}}
\end{center}
 \vspace{-0.5cm}
\end{table}

\captionsetup[figure]{name=Extended Data Figure}
\setcounter{figure}{0}

\begin{figure}[ht]
\center
 \includegraphics[width=0.6\columnwidth]{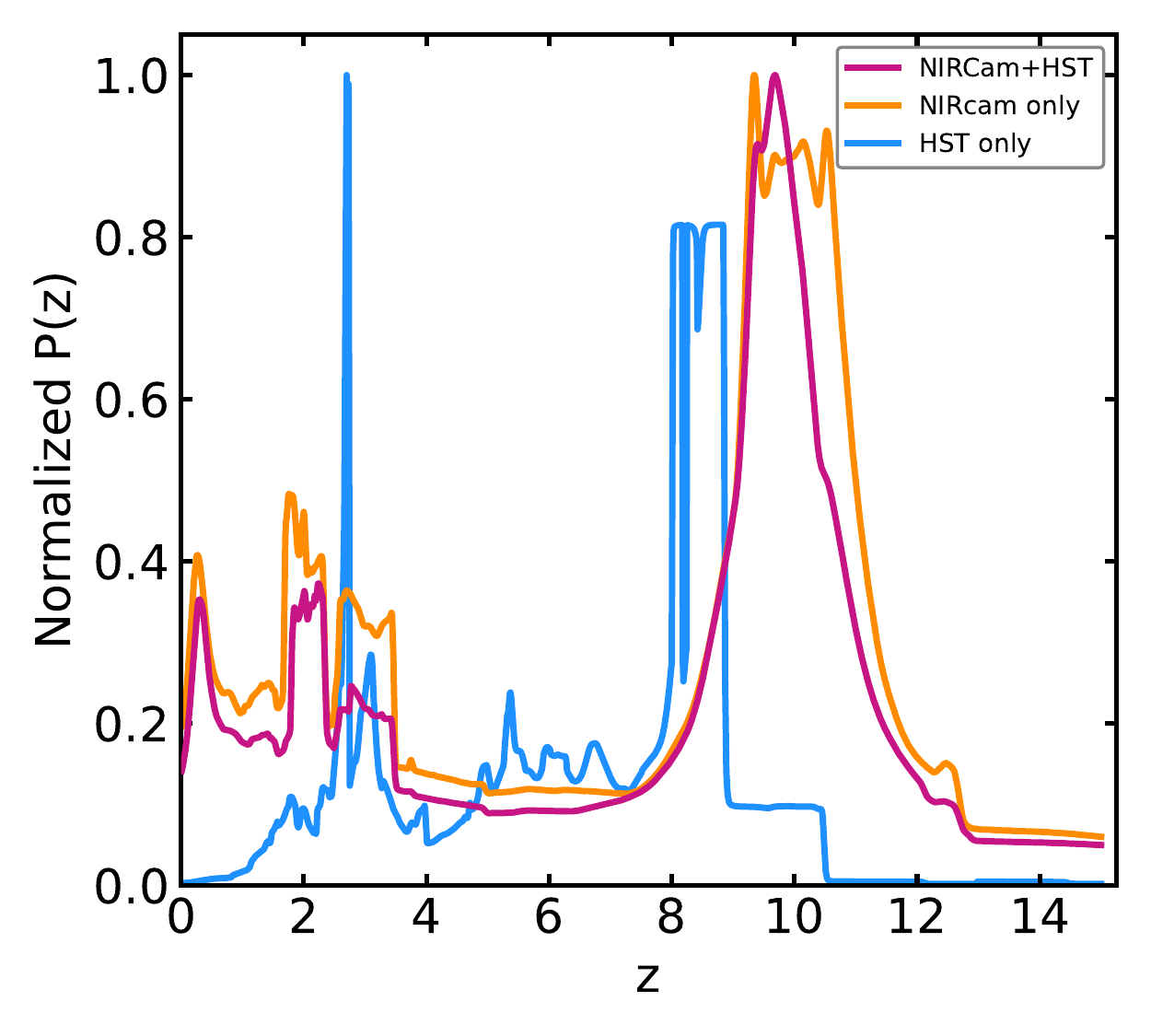} \\
 {\small \caption{\textbf{The $P(z)$ of JD1 from \texttt{EAzY} fits to a variety of photometry}. Each $P(z)$ is constructed through \texttt{EAzY} modelling of three sets of photometric data points. Blue adopts \textit{HST} photometry only\cite{zitrin14}, orange adopts NIRCam photometry only, and red illustrates results for the combined data set. The \textit{HST} photometry provides excellent constraining power blueward of the $J$-band, but lacks the wavelength coverage in the IR that NIRCam provides to exclude low-$z$ interlopers. As such, even using NIRCam data alone provides far better constraints for $z>9$ selections. NB: The photo-$z$ estimated by [\!\!\citenum{zitrin14}] also incorporates \textit{Spitzer/IRAC} photometry, not used here.}\label{fig:pdf}}
\end{figure}


\end{document}